# Particle size effect on the Langmuir-Hinshelwood barrier for CO oxidation on regular arrays of Pd clusters supported on ultrathin alumina films

Georges Sitja, Héloïse Tissot and Claude R. Henry

*Aix-Marseille Université, CNRS, CINaM, F-13000 Marseille, France*

Corresponding author email address : henry@cinam.univ-mrs.fr

**Abstract**

The Langmuir-Hinshelwood barrier ($E_{LH}$) and the pre-exponential factor ($v_{LH}$) for CO oxidation have been measured at high temperature on hexagonal arrays of Pd clusters supported on an ultrathin alumina film on $Ni_3Al(111)$. The Pd clusters have a sharp size distribution and the mean sizes are: 174±13, 360±19 and 768±28 atoms. $E_{LH}$ and $v_{LH}$ are determined from the initial reaction rate of a CO molecular beam with a saturation layer of adsorbed oxygen on the Pd clusters, measured at different temperatures (493≤T (K) ≤613). The largest particles (3.5 nm) give values of $E_{LH}$ and $v_{LH}$ similar to those measured on Pd (111) [2]. However, smaller particles (2.7 and 2.1 nm) show very different behavior. The origin of this size effect is discussed in terms of variation of the electronic structure and of the atomic structure of the Pd clusters.

## I. Introduction

The oxidation of CO on transition metals is a prototypical reaction in heterogeneous catalysis [1]. The mechanism of this reaction has been definitely established by the group of Ertl on Pd (111) [2] and later on Pt (111) [3] using the modulated molecular beam reactive scattering technique (MBRS). The reaction follows the Langmuir-Hinshelwood mechanism where an adsorbed CO molecule and an adsorbed oxygen atom, provided by the dissociation of a dioxygen molecule, react to form a $CO_2$ molecule. This reaction is considered as 'structure insensitive' that means that it does not depend on the structure of the crystal surface [4]. Thus, it was assumed that it did not depend on particle size in the case of supported catalysts [5]. This assumption was verified on Pd particles of different mean sizes supported on α-alumina at 445 K (CO rich regime). However, it was observed that at 518 K (O-rich regime), the turnover frequency (TOF) increased when the particle size decreased [5]. This effect was later explained by the reverse spillover (i.e. the capture of CO molecules physisorbed on the alumina support by the Pd particles) [6, 7] which has nothing to do with the concept of 'structure sensitivity'.

Nevertheless, in these early studies on supported model catalysts, the size dispersion of metal particles was a major problem impeding a thorough investigation into particle size effects in catalysis, in particular for small cluster size [8]. A good way to circumvent the problem of size dispersion is to grow the nanoparticles on a template [9]. The template must present a regular array of nucleation centers where nucleation occurs rapidly, and allow a uniform growth leading to metal clusters arrays with a very sharp size distribution. Up to now, three types of templates have been used to prepare supported metal catalysts: graphene monolayer [10], h-BN monolayer [11] grown on a metal substrate and alumina ultrathin film grown on a $Ni_3Al$ (111) surface [12]. Ultrathin alumina film on $Ni_3Al$ (111) appears to be the best template to prepare supported model catalysts because the organization remains up to 600 K and no cluster coalescence nor intercalation occurs during gas exposure [13]. Therefore, size controlled metal clusters grown on ultrathin alumina film on $Ni_3Al$ (111) constitute a perfect model system to perform systematic studies and gain knowledge on small clusters behavior.

Using this template the adsorption energy of CO has been measured on arrays of Pd clusters from 5 atoms to a mean size of 4 nm [14]. The adsorption energy decreases monotonically with clusters size down to roughly 2.5 nm. For smaller clusters, the adsorption energy varies non-monotonically with size reflecting the non-metallic electronic structure of the clusters [14]. Later on, the CO oxidation was studied by MBRS on these arrays of Pd clusters [15]. In this study, the effect of the reverse spillover on the CO adsorption and oxidation has been studied and the steady state CO oxidation rate has been measured as a function of temperature. After correction from the reverse spillover, the TOF of $CO_2$ production was measured for a cluster size of 181± 13 atoms. At high temperature (T ≥ 500K) the TOF was close to the one obtained on Pd (111) in similar experimental conditions [2], while at low temperature (T< 500K) the reactivity was definitely higher for Pd clusters than for Pd (111). The increase of reactivity at low temperature for the Pd clusters was explained by the decrease of CO adsorption energy that reduces CO poisoning [15].

The Langmuir-Hinshelwood barrier ($E_{LH}$) for CO oxidation reaction has been measured by Engel and Ertl on Pd (111) [2]. From transient experiments, they have found at high temperature (T≥ 500 K) $E_{LH}$ =112.9 ± 8 kJ/mole in agreement with measurements by MBRS ($E_{LH}$ =104.5 ± 8 kJ/Mol.). This value is only valid for a very low CO coverage and an oxygen saturation coverage since $E_{LH}$ depends on both CO and oxygen coverages (this crucial point in the determination of $E_{LH}$ will be discussed in the paper). At lower temperature (T < 500 K), $E_{LH}$ decreased and was found equal to 58.5± 8 kJ/mole due to an increase of CO coverage [2].

In this paper we will focus on extending these investigations to nanoparticles in order to have a better understanding of the particle size effect during CO oxidation. The Langmuir-Hinshelwood barrier ( $E_{LH}$ ) and the pre-exponential factor ($v_{LH}$) will be determined using MBRS for different sizes of Pd nanoparticles. Methods to obtain these quantities will be described and results will be discussed and compared with the literature data.

## II. Experimental

Details on the alumina films preparation have been described in a previous paper [16]. Briefly, the alumina film is formed by oxidation of a $Ni_3Al$ (111) surface exposed to 40 L of oxygen at 1000 K in a background oxygen pressure of $5 \times 10^{-8}$ mbar. Pd clusters are grown at 373 K on the alumina film by using a liquid nitrogen-cooled Knudsen cell. The Pd flux is calibrated *in situ* by a quartz crystal micro-balance. From previous in situ STM and GISAXS measurements [13] we know that the Pd clusters form a hexagonal lattice with a lattice parameter of 4.1 nm. The arrays of Pd clusters are perfectly reproducible and the mean cluster size depends only of the Pd deposition time. A typical STM picture of an array of Pd clusters is shown on Figure 1. The average number of atoms in the clusters is obtained by dividing the amount of deposited Pd atoms by the density of clusters which is $6.5 \times 10^{12}$ $cm^{-2}$. We have previously shown that the size dispersion of a cluster containing N atoms is $1/\sqrt{N}$ [14]. The

CO oxidation experiments are performed *in situ* in the UHV chamber where the clusters are grown. A molecular beam of CO is generated by supersonic expansion and modulated as square pulses which directly strike the sample in a single collision (i.e. no interaction with the chamber walls) [17]. The intensity of the CO beam at the sample level is equal to $2 \times 10^{13}$ molecules.$cm^{-2}$ which is equivalent to an isotropic CO pressure of $6.8 \times 10^{-8}$ mbar. The oxygen pressure is isotropic and regulated through a leak valve. The desorbed CO and the $CO_2$ product are measured with a mass spectrometer.

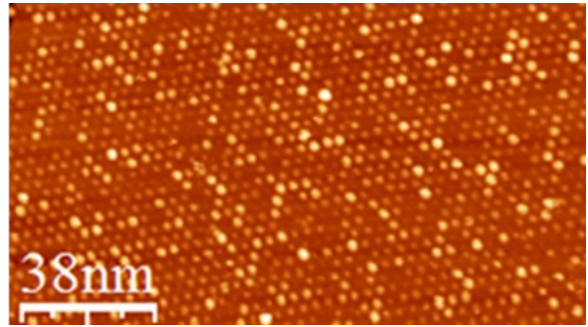

*Figure 1: STM image of 0.15 Pd deposited on alumina/Ni3Al (111) showing the hexagonal organization of the clusters.*

For the measurement of $CO_2$ transients two methods have been used. The critical point here is that $E_{LH}$ depends on both CO and oxygen coverages; therefore, the experiment must be performed in a situation where both these coverages are easy to determine. For the oxygen coverage, conditions were chosen in order to have a saturated surface for which the coverage is known and for CO, coverage is determined using the calibrated beam flux as will be described later.

**First method:** The Pd clusters are exposed at an oxygen pressure of $6 \times 10^{-8}$ mbar up to the saturation of the oxygen adsorbed layer. For Pd clusters, we have previously shown that the saturation coverage is $\Theta_{sat} = 0.4$ [18]. The oxygen pressure is pumped down and the sample is exposed to a pulse of CO. Then, the $CO_2$ produced by the reaction of CO with the oxygen adlayer is recorded until the complete consumption of adsorbed oxygen on the Pd particles (see two examples of recorded $CO_2$ pulses obtained at two different temperatures on Fig. 2). During the $CO_2$ pulse, the coverages of CO ($\Theta_{CO}$) and oxygen ($\Theta_O$) vary. In order to have a well-defined value of $\Theta_O$ and a low value of $\Theta_{CO}$, we take the initial value of the $CO_2$ desorbed flux which corresponds to a time of 0.1 second. In these condi-tions, we consider that the oxygen coverage is still the saturation value. The CO coverage being too small to be measured; we assume that it is equal to the equilibrium coverage ($\Theta_{eq}$) that we calculate from eq.1. In fact this value corresponds to a pure adsorption/desorption phenomenon and it is an upper value of the true coverage during the reaction. This approximation was also used by Engel and Ertl [2] to derive the value of $E_{LH}$ at high temperature. To justify this approximation we can compare the adsorption rate with the reaction rate. The adsorption rate is always much larger than the reaction rate in the condition of the measurements (first point of the transient $CO_2$ production).

$$\Theta_{eq} = \alpha_g J_{CO} \tau / N_{Pd} \qquad (1)$$

$\alpha_g$ is the global adsorption coverage that represents the proportion of the impinging CO flux ($J_{CO}$) that is chemisorbed on the Pd particles. $\alpha_g$ is calculated from the analytic model developed in order to take quantitatively into account the effect of the reverse spillover [19] using the parameters $\alpha$ (the adsorption probability of CO on the alumina film) and $E_a$-$E_d$ (the difference between the adsorption energy and the diffusion energy of a CO molecule physisorbed on the alumina film) being equal to 0.4 and 0.15 eV, respectively. These values were experimentally determined for regular arrays of Pd clusters on ultrathin alumina films on $Ni_3Al$ (111) [15]. $\tau$ is the lifetime of a CO molecule adsorbed on a Pd cluster which is calculated from eq. 2. $N_{Pd}$ is the number of Pd surface atoms into the clusters supported on a unit surface of 1 cm². It is calculated knowing the radius of the clusters and assuming an hemispherical shape [13].

$$\tau = \nu^{-1} \exp (E_{ad}^{CO}/RT) \qquad (2)$$

$E_{ad}^{CO}$ is the adsorption energy of CO on the Pd clusters and $\nu$ is the frequency factor; these two parameters have been measured by MBRS for different cluster sizes [14] and verified for each sample before and after CO oxidation experiments. The fact that there is no evolution of the adsorption energy of CO after CO oxidation shows that the reaction has not induced an irreversible change of the particle morphology. Values of $E_{ad}^{CO}$ and of the frequency factor are indicated in Table 1.

**Second method:** In the second method, we keep the oxygen pressure constant during the entire measurement cycle in order to ascertain the saturation oxygen coverage on the Pd particles at the beginning of the CO pulse avoiding the pumping time of the oxygen background pressure as in the first method. The CO beam is opened during 2 seconds and kept closed during 30 seconds in order to replenish the oxygen layer on the Pd clusters before a new cycle. We also increase the time resolution in order to have a better definition of the initial oxygen and CO coverages (just at the opening of the CO beam) but the price to pay is a decrease of the signal/noise ratio; then we average the signal over 10 cycles to overcome the deterioration of the signal/noise ratio. The measurements for clusters of 174 atoms where done in these conditions. After this series of measurements we have modified the experimental setup in order to increase significantly (by a factor of about 8) the signal to noise ratio in the measurements (basically the distance between the sample and the nose of the mass spectrometer has been strongly reduced). Figure 3 shows examples of CO$_2$ transients using the second method after the modification of the experimental setup. To determine accurately the initial value of the CO$_2$ signal which corresponds to a time of 0.02 s, we fitted the CO$_2$ signal by an exponential function. For this short delay, the CO coverage has not necessary reached its equilibrium value, therefore we explicitly calculate $\Theta_{CO}$ with equation 3.

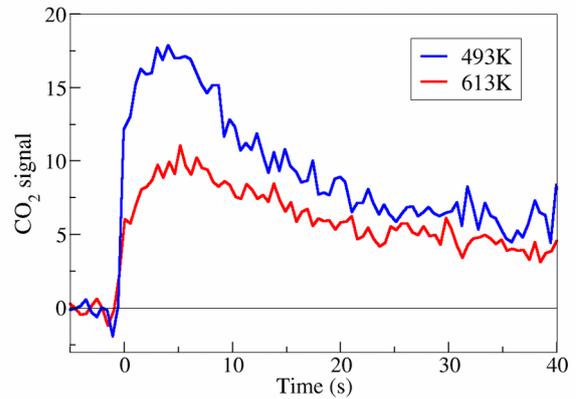

*Figure 2: Examples of $CO_2$ transients obtained with the first method at 493 and 613 K on an array of Pd clusters containing 174±13 atoms.*

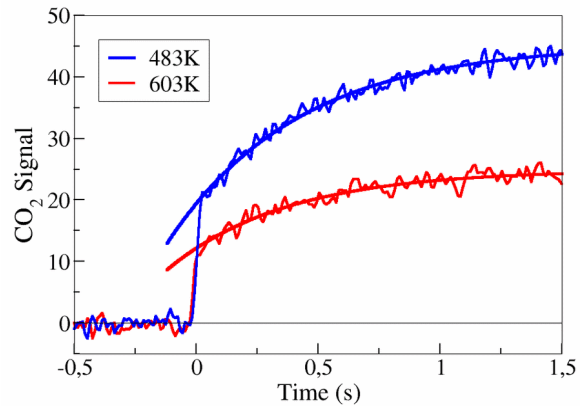

*Figure 3: Examples of $CO_2$ transients obtained with the second method at 483 and 603 K on an array of Pd clusters containing 768±28 atoms. The $CO_2$ signal is fitted by an exponential function (continuous curve).*

$$\Theta_{CO} \approx \alpha_g J_{CO} \tau\ [1-\exp(-t/\tau)]/N_{pd} \quad (3)$$

## III. Results

The two methods have been first tested on Pd clusters containing 174±13 atoms that correspond to a diameter of 2.1 nm assuming a hemispherical shape which is justified by the aspect ratio close to 1/2 measured previously [13]. The CO$_2$ transient have been measured at T= 493, 553, 583 and 613 K. The equilibrium CO coverage varies between 2.5x10$^{-4}$ at 613K to 1.4x10$^{-3}$ at 493 K. The initial value of the CO$_2$ signal decreases when the substrate temperature increases (see fig. 2). The reaction rate (per atomic site) can be written as:

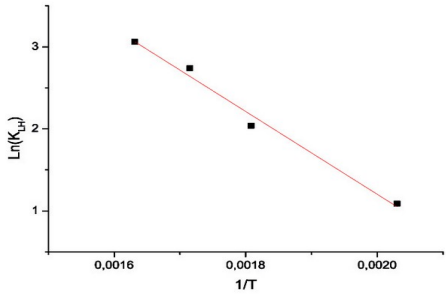

*Figure 4: Arrhenius plot of $K_{LH}$ obtained with the first method for clusters of 174 ±13 atoms. The red line is the linear fit of experimental data. The slope gives $E_{LH}$ = 41.95 ±3.3 kJ/Mole and the intercept gives $\nu_{LH}$ = 8.02x10$^4$ s-1.*

$$r_{CO2} = k_{LH}\ \Theta_o\ \Theta_{CO} \quad (4)$$

with

$$k_{LH} = \nu_{LH}\ \exp(-E_{LH}/RT) \quad (5)$$

$\nu_{LH}$ is the pre-exponential factor and $E_{LH}$ is the Langmuir-Hinshelwood barrier. Then, plotting ln($k_{LH}$) vs 1/T should give a straight line where the slope is equal to $E_{LH}$/R and the intercept is ln($\nu_{LH}$). In order to obtain the correct value of the pre-factor, the CO$_2$ signal must be properly calibrated. For this calibration, we recorded the maximum of the CO desorbed flux which is equal, in the absence of oxygen pressure, to the incident

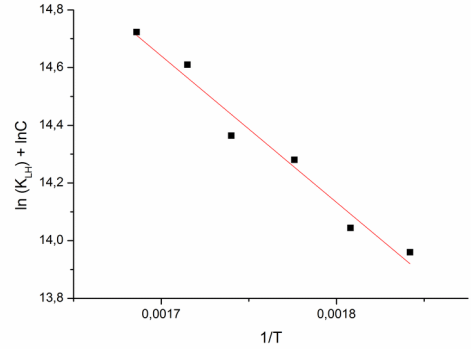

*Figure 5: Arrhenius plot of $K_{LH}$ obtained with the second method for clusters of 360 ±19 atoms. The red line is the linear fit of experimental data. The slope gives $E_{LH}$ = 42.2 ±3.5 kJ/Mole, $\nu_{LH}$ could not be measured because the calibration constant C of the CO2 flux was not determined for these measurements.*

flux (2x10$^{13}$ molecules.cm$^{-2}$.s$^{-1}$). Then the CO$_2$ signal is calibrated during CO oxidation knowing that the missing CO at steady state corresponds to CO$_2$ formation. Figure 4 displays the Arrhenius plot for the Pd clusters containing 174 atoms. The

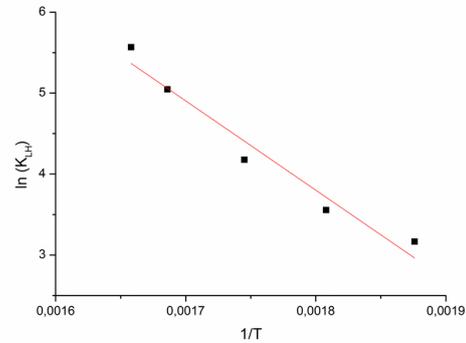

*Figure 6: Arrhenius plot of $K_{LH}$ obtained with the second method for clusters of 768 ±28 atoms. The red line is the linear fit of experimental data. The slope gives $E_{LH}$ = 91.66 ± 8.8 kJ/Mole and the intercept gives $\nu_{LH}$ =1.9x10$^{10}$ s-1.*

$E_{LH}$ value is 41.95 ±3.3 kJ/mole and the pre-factor 8x10$^4$ s$^{-1}$ (see table 1). With the second method time resolution is improved and the accuracy in the $E_{LH}$ and $\nu_{LH}$ measurements is expected to be better. However, in increasing the time resolution the signal to noise ratio decreases strongly leading to a large error in the $E_{LH}$ value. We obtain (see table 1): $E_{LH}$ = 56.7±16 kJ/mole and $\nu_{LH}$ = 1.5x10$^6$ s$^{-1}$. The value of $E_{LH}$ with the second method is larger than those obtained with the first method but it is definitively different from those obtained on Pd (111) (see table 1). Taking into account the large error bar for the value of $E_{LH}$ given with the second method it remains compatible with those obtained with the first method. Later on, similar experiments using only the second method were performed for larger particle sizes: 360±19 atoms and 768±28 atoms after a modification of the experimental setup (see Experimental section) in order to improve notably (by a factor of 8) the signal/noise ratio. For clusters of 360 atoms (D=2.7 nm), we made measurements between 543 and 593 K which correspond tto 3x10$^{-5}$≤Θ$_{CO}$ ≤1x10$^{-4}$. The Arrhenius plot (see figure 5) gives $E_{LH}$ = 42.2 ±3.5 kJ/mole for 360 atoms clusters which is close to the values obtained for the smaller clusters. It has not been possible to determine the value of the pre-factor $\nu_{LH}$ for this cluster size because the CO$_2$ signal was not calibrated.

| i (atoms) | D (nm) | method | $E_{ad}^{CO}$ (kJ/mole) | $\nu_{CO}$ (s$^{-1}$) | $E_{LH}$ (kJ/mole) | $\nu_{LH}$ (s$^{-1}$) |
|---|---|---|---|---|---|---|
| 174±13 | 2.1 | 1st | 93.3 | 2.6x10$^9$ | 41.95±3.3 | 8.02x10$^4$ |
| 174±13 | 2.1 | 2nd | | | 56.72±16 | 1.5x10$^6$ |
| 360±19 | 2.7 | 2nd | 91.1 | 2.4x10$^{10}$ | 42.2±3.5 | - |
| 768±28 | 3.5 | 2nd | 132.5 | 1.5x10$^{14}$ | 91.66±8.8 | 1.9x10$^{10}$ |

*Table 1*: Langmuir-Hinshelwood energy barrier and pre-exponential factor measured at high temperature for regular arrays of Pd clusters containing i atoms and having a diameter D. The adsorption energy ($E_{ad}^{CO}$) and the frequency factor ($\nu_{CO}$) for CO adsorbed molecules on the different clusters are also indicated.

The largest particles we have studied contained 768±28 atoms which correspond to a diameter of 3.5 nm. Figure 6 displays the Arrhenius plot of the measurements, using the second method, obtained between 533 and 603 K which correspond to CO coverage between 1.7x10$^{-4}$ and 2.0x10$^{-5}$. However, it is interesting to note that below 523 K, data are no longer on the straight line obtained at higher temperature indicating that the influence of the CO coverage becomes important below this temperature. The same trend was already observed on Pd (111) below 500K [2]. The value of $E_{LH}$ (91.66 ±8.8 kJ/mole) and of the pre-factor (1.9x10$^{10}$ s$^{-1}$) obtained from the Arrhenius plot are much larger than for smaller clusters and close to those previously measured on Pd (111) [2] (see table 1).

## IV. Discussion

Table 2 displays the values obtained experimentally or theoretically, in previous studies, for the Langmuir-Hinshelwood barrier and the pre-factor for CO oxidation on Pd extended surfaces or supported particles.

On Pd (111) Engel and Ertl [2] have measured $E_{LH}$ values of 104 kJ/mole using the CO$_2$ transient method and 113 kJ/mole using the MBRS technique at temperatures higher than 500 K corresponding to a low coverage of CO and an oxygen coverage close to saturation (0.25). These values of $E_{LH}$ are in rather good agreement with ab-initio calculations [20, 21] for CO oxidation on Pd (111). They also found that the Langmuir-Hinshelwood rate constant depends on CO and O coverages [2]. At low temperature (i.e. high CO coverage), they find a much lower value for $E_{LH}$ (58.5 kJ/mole). In the present study we find for Pd clusters containing 768 atoms (D=3.5 nm) $E_{LH}$ = 91.7 kJ/mole which is close to the value obtained by Engel and Ertl for Pd (111). Moreover, the pre-exponential factors have similar values. Then, we can safely conclude that down to this size there is no size effect. Two previous studies have been devoted to the experimental determination of the Langmuir-Hinshelwood barrier for CO oxidation on supported Pd particles. In the first one, by Matolin group [22], $E_{LH}$ has been measured by the transient CO$_2$ experiments at low temperature (410 K) for Pd (111) and for Pd particles on alumina with diameters of 27 and 2.5 nm. For Pd (111) they obtain a value of $E_{LH}$ rather close to the one measured by Engel and Ertl at low temperature on Pd (111). However, for Pd particles the Langmuir-Hinshelwood energy decreases with cluster size for the two studied sizes. The authors concluded on the existence on size effect already observable for particles of 27 nm. In the second study performed in

| System | Method | T (K) | $\Theta_o$ | $\Theta_{co}$ | $E_{LH}$ (kJ/mole) | $\nu_{LH}$ (s$^{-1}$) | reference |
|---|---|---|---|---|---|---|---|
| Pd(111) | transient | >500 | 0.25 | low ($\Theta_{eq}$) | 112.9 | $\approx 1 \times 10^{11}$ | |
| | | <500 | 0.25 | >0.02 | 58.5 | $\approx 10^6$ | 2 |
| | MBRS | ≥550 | >0.1 | low | 104.5 | | |
| Pd(111) | Theory | | 0.25 | low | 134 | $2 \times 10^{13}$ | 20 |
| Pd (100) | Theory | | 0.25 | low | 101 | $2 \times 10^{12}$ | 20 |
| Pd(111) | Theory | | low | low | 108 | $10^{13}$ | 21 |
| Pd(111) | transient | 410 | ≥0.25 | ≤0.06 | 64/45 | - | 22 |
| Pd/Al$_2$O$_3$ D=27nm | transient | 410 | ≥0.25 | ≤0.06 | 45/32 | - | 22 |
| Pd/Al$_2$O$_3$ D=2.5nm | transient | 410 | ≥0.25 | ≤0.06 | 20/19 | - | 22 |
| Pd/Al$_2$O$_3$/NiAl(110) D=5-6 nm | transient | 350-450 | ≥0.25 | 0.07-0.25 | 57-62 | - | 23 |
| Pd$_{30}$/MgO | Theory | | low | low | 82.6/111.5 | - | 27 |
| | | | 0.4 | 0.5 | 38.5/63.4 | - | |

*Table 2: Previous experimental or theoretical determinations of the Langmuir-Hinshelwood barrier and pre-exponential factor for CO oxidation on Pd surfaces or on supported Pd clusters.*

Freund's group [23], $E_{LH}$ has been measured by the $CO_2$ transient method at low temperature (350 – 450 K) for Pd particles with a diameter of 5 to 6 nm supported on an alumina film on NiAl (110). They found $E_{LH}$ values between 57 to 62 kJ/mole. As this value is comparable to the one obtained on Pd (111), at low temperature, they concluded that down to 5 nm there is no size effect. This conclusion is in agreement with the present measurements (at high temperature) and confirms the absence size effect down to these sizes. Our results allow us to extend this conclusion down to particles with a diameter of 3.5 nm. For smaller particles, at high temperature (T>500K) for Pd particles of 2.7 nm (360 atoms,) we observe an important drop of $E_{LH}$ (roughly half value) which is confirmed at D = 2.1 nm (177 atoms). This strong decrease of $E_{LH}$ is accompanied with a large decrease of the pre-exponential factor. These observations suggest a strong size effect in this size range (D ≤ 2.7 nm). The size effect observed here is not fully in agreement with the results of Matolin et al [22] besides the fact that their measurements were made at low temperature, they also observe a large decrease of $E_{LH}$ already for particles of 27 nm in disagreement with our measurements and those of Freund's group [23]. This discrepancy could be due to the large size dispersion obtained with the growth method using a thick polycrystalline alumina substrate in the work of Matolin et al. [22].

What can be the origin of the observed size effect? One can suggest the increase of the oxygen coverage and the presence of subsurface oxygen on small Pd particles. However, this effect is already present in the work of Freund's group [23] for particles of 5-6 nm for which no size effect exist. Another explanation could be the variation of the adsorption energy of CO with cluster size as already studied by Sitja et al. on the same system [14]. Two regimes were observed: from ~2.5 nm to 6 nm the adsorption energy increases continuously with cluster size (scalable regime) in agreement with previous measurements on Pd clusters supported on a $Fe_3O_4$ film on Pt (111) [24], while below 2.5 nm the adsorption energy depends on the number of atoms in the cluster in a non-monotonic way. We have measured the steady state reaction rate for CO oxidation on Pd clusters containing 181 atoms (2.1 nm), a higher activity compared to Pd (111) was observed at low temperature (T<500K) but at high temperature the activity was the same [15]. Thus, at first sight, one could conclude that the change of the adsorption energy is not at the origin of the size effect observed here (i.e. for 171 atoms). However, at high temperature the reaction is limited by the rapid desorption of CO, the decrease of the adsorption energy of CO would decrease the reaction rate. Meanwhile, the decrease of $E_{LH}$ would increase the reaction rate, the fact that the TOF for Pd clusters of 2.1 nm (i=174 atoms) is close to that one measured on Pd (111) would be due to a compensation of the two opposite effects. For larger cluster (i=360 atoms) both the value of $E_{LH}$ and the adsorption energy of CO have strongly decreased compared to large Pd particles (i= 768 atoms) or Pd (111) [2]. Thus it is possible that the decrease of $E_{LH}$ and the decrease of the CO adsorption energy have the same origin but the small range of size studied here does not allow to definitely conclude on the origin of these decays. An alternative explanation of the decay of $E_{LH}$ when particle size decreases could be the increase of the proportion of edges sites at low sizes. If we consider our Pd clusters as half truncated octahedra (supported by STM , GISAXS and grazing incidence X-ray diffraction measurements [9]) the ratio of the number of edge atoms to the number of face atoms for the studied sizes is 0.25, 0.32 and 0.50 for 768 atoms, 360 atoms and 174 atoms respectively. Thus , between 360 atoms and 174 atoms the proportion of edge atoms increases two times more (+56%) than between 768 to 360 atoms (+28%). Meanwhile, between 360 and 174 atoms $E_{LH}$ is more or less constant and between 768 and 360 atoms $E_{LH}$ decreases almost by a factor of 2. Moreover between Pd (111) and clusters of 768 atoms the proportion of edge atoms increases from almost zero to 0.25 while $E_{LH}$ is nearly constant. Then we can hardly tell that there is correlation between the proportion of step edges and the decrease of $E_{LH}$ for clusters having the shape of an half truncated octahedron.

A further explanation could be the change of the electronic structure of the cluster between the scalable and the non-scalable regime. Indeed, it is known from photoemission studies on free metal clusters (Cu, Ag, Au) that below ~100 atoms, the electronic structure is constituted of discrete energy levels while for larger clusters it is made of continuous bands [25]. These two regimes are responsible for significant changes in the chemical properties as showed for CO adsorption energy on Pd clusters where the transition between the two regimes is observed around 100 to 200 atoms [14]. In the case of Pt, recent ab initio calculations have shown that the transition between the two regimes is just below 147 atoms [26]. Without further theoretical calculations, it is difficult to attribute the drastic change of $E_{LH}$ for Pd clusters containing 360 atoms to the change of the electronic structure. Ab initio calculations have been performed for CO oxidation on 30 atoms Pd clusters supported on MgO (100) [27]. The Pd cluster is a pyramid exposing (111) facets. At low temperature and high CO and O coverages, the activation barrier for the reaction varies between 38.5 to 63.4 kJ/mole while at high temperature (low coverage of CO) the activation energy is between 82.6 to 111.5 kJ/mole, depending on the configuration of the reactants. These values at low and high temperature are in fair agreement with the experimental results on Pd (111) in similar conditions. Thus, from these calculations we can conclude in the absence of a size effect for the 30 atoms Pd clusters, at least if the Pd clusters keep the same structure and morphology during the reaction.

Indeed, another explanation of the observed size effect could be a reversible change of morphology or roughness of the Pd clusters during the reaction. In a recent work in ETEM (Environmental Transmission Electron Microscopy) [28], it has been shown that Au particles containing about 1000 atoms (4 nm) supported on $CeO_2$ are single crystals with well-developed low index facets. This shape is stable in vacuum and during CO oxidation at RT at a pressure of $10^{-4}$ mbar, only an increase of the proportion of (100) facets is observed. Particles of 100 to 200 atoms (<2nm) have also a facetted shape in vacuum but, during CO oxidation, particles appear hemispherical and lose their fcc structure. Single layer clusters with about 40 atoms become 3D with no crystalline structure during CO oxidation. To the best of our knowledge, no such experiments have been yet performed for supported Pd particles smaller than 200 atoms. However, if the same size dependent dynamical restructuring during CO oxidation would exist, one could expect important changes in the energetic of the reaction that could explain the drastic decrease of $E_{LH}$ from a particle size of 360 atoms.

## V. Summary

We have studied the size dependence of the Langmuir-Hinshelwood barrier ($E_{LH}$) for CO oxidation on Pd clusters supported on alumina film. The growth of the clusters on a template formed by a nanostructured ultrathin alumina film on $Ni_3Al$ (111) allows the formation of a hexagonal array of clusters with a sharp size distribution and a high number density ($6.5 \times 10^{12}$ cm$^{-2}$). $E_{LH}$ is measured by transient $CO_2$ production in conditions for which the energy barrier is constant consisting of a saturation coverage of oxygen, a very low coverage of CO and a high temperature (493-613K). For Pd clusters containing 768±28 atoms (D=3.5 nm), $E_{LH}$ and the pre-exponential factor have similar values than those obtained on Pd (111). For Pd clusters of 360 and 174 atoms $E_{LH}$ decreases drastically (divided by a factor of two). Different possible origins of this size effect arising at a size of 360 atoms are discussed and a dynamical structural evolution of the cluster shape/structure, during the reaction, seems the most probable explanation.

# References

[1] H.J. Freund, G. Meijer, M. Scheffler, R. Schlögl and M. Wolf, Angew. Chem., Int. Ed. **50**, 10064 (2011)

[2] T. Engel and G. Ertl, J. Chem. Phys. **69**, 1267 (1978)

[3] C.T. Campbell, G. Ertl, H. Kuipers and J. Segner, J. Chem. Phys. **73**, 5862 (1980)

[4] T. Engel and G. Ertl, Adv. Catalysis **28**, 1 (1979)

[5] S. Ladas, H. Poppa and M. Boudart, Surf. Sci. **102**, 151 (1981)

[6] F. Rumpf, H. Poppa and M. Boudart, Langmuir **4**, 722 (1988)

[7] C.R. Henry, Surf. Sci. **223**, 519 (1989)

[8] C.R. Henry, Surf. Sci. Rep. **31**, 235 (1998)

[9] C.R. Henry, Catal. Lett. **145**, 731 (2015)

[10] T. Gerber, J. Knudsen, P.J. Feibelman, E. Granas, P. Stratmann, K. Schulte, J.N. Andersen and T. Michely, ACS Nano **7**, 2020 (2013)

[11] M. Will, N. Atodiresei, V. Casiuc, P. Valerius, C. Herbig and T. Michely, ACS Nano **12**, 6871 (2018)

[12] S. Degen, C. Becker and K. Wandelt, Faraday Discuss. 125, 343 (2004)

[13] M. Marsault, G. Sitja and C.R. Henry, Phys. Chem. Chem. Phys. 16, 26458 (2014)

[14] G. Sitja, S. Le Moal, M. Marsault, G. Hamm, F. Leroy and C.R. Henry, Nano Lett. **13**, 1977 (2013)

[15] G. Sitja and C.R. Henry, J. Phys. Chem. C **121**, 10706 (2017)

[16] G. Hamm, C. Barth, C. Becker, K. Wandelt and C.R. Henry, Phys. Rev. Lett. 97, 126106 (2006)

[17] C.R. Henry, in *The Chemical Physics of Solid Surfaces*, Vol. 11: *Surface Dynamics*, D.P. Woodruff Ed.; Elsevier, 2003, P. 247

[18] L. Piccolo, C. Becker and C.R. Henry, Appl. Surf. Sci. **164**, 156 (2000)

[19] C.R. Henry, C. Chapon and C. Duriez, J. Chem. Phys. **95**, 700 (1991)

[20] A. Eichler, Surf. Sci. **498**, 314 (2002)

[21] S. Piccinin and M. Stamatakis, ACS Catal. **4**, 2143 (2014)

[22] I. Stara, V. Nehasil and V. Matolin, Surf. Sci. **331-333**, 173 (1995)

[23] I. Meusel, J. Hoffmann, J. Hartmann, M. Heemeier, M. Bäumer, J. Libuda and H.J. Freund, Catal. Lett. **71**, 5 (2001)

[24] J.M. Flores-Camacho, J.H. Fischer-Wolfarth, M. Peter, C.T. Campbell, S. Schauermann, H.J. Freund, Phys. Chem. Chem. Phys. **13**, 16800 (2011)

[25] K.J. Taylor, C.L. Pettiette-Hall, O. Cheshnowsky and R.E. Smalley, J. Chem. Phys. **96**, 3319 (1992)

[26] L. Li, A.H. Larsen, N.A. Romero, V.A. Morozov, C. Glinsvad, E. Abild-Pedersen, J. Greeley, K.W. Jacobsen and J.K. Norskov, J. Phys. Chem. Lett. **4**, 222 (2013)

[27] B. Yoon, U. Landman, V. Habibpour, C. Harding, S. Kunz, U. Heiz, M. Moseler and M. Walter; J. Phys. Chem. C **116**, 9594 (2012)

[28] Y. He, J.C. Liu, L. Luo, Y.G. Wang, J. Zhu, Y. Du, J. Li, S.X. Mao and C. Wang, PNAS **115**, 7700 (2018)